\documentclass[aps,twocolumn,prl]{revtex4}
\usepackage{exscale}
\usepackage{graphicx}
\usepackage{amsmath}
\usepackage{latexsym}
\usepackage{amsfonts}
\usepackage{amssymb}

\def\pmx{\begin{pmatrix}}
\def\emx{\end{pmatrix}}

\newcommand{\ket}[1]{|#1\rangle}

\begin{document} 

\title{Multipartite entanglement in quantum algorithms}

\author{
D. Bru{\ss}$^1$ and C. Macchiavello$^2$}

\affiliation{$^1$Institut f{\"u}r Theoretische Physik III, 
Heinrich-Heine-Universit{\"a}t D{\"u}sseldorf, D-40225 D{\"u}sseldorf, 
Germany \\
$^2$Dipartimento di Fisica ``A. Volta" and INFN-Sezione di Pavia,
Via Bassi 6, 27100 Pavia, Italy}

\begin{abstract}

We investigate the entanglement features of the quantum states employed in 
quantum algorithms. In particular, we analyse the multipartite entanglement
properties in the Deutsch-Jozsa, Grover and Simon algorithms.
Our results show that for these algorithms most instances
involve multipartite entanglement. 
\end{abstract}

\date{\today}

\maketitle

Entanglement is a major resource in quantum information processing. However, 
its role in achieving the quantum computational speedup in the currently known
quantum algorithms is not yet completely clear and has been a highly debated 
question since the advent of quantum computation. In particular, the role of
genuine {\em multipartite} entanglement has not been elucidated thoroughly.
It was shown that in  Shor's algorithm  multipartite entanglement is needed
to achieve exponential computational speedup with quantum resources \cite{jl}.
In this work we analyse other known quantum algorithms, namely 
Deutsch-Jozsa, Grover and Simon's, and assess the 
multipartite entanglement properties of the pure quantum states employed. 

Let us first focus on the Deutsch-Jozsa \cite{dj} and Grover \cite{grover} 
algorithms, following the formulation given in \cite{cemm}.
The $n$-qubit states that occur in these algorithms are of the form

\begin{equation}
\ket{\psi_f}\equiv \frac{1}{\sqrt{2^n}}\sum_{x=0}^{2^n-1}(-1)^{f(x)}\ket{x}\;,
\label{psif}
\end{equation}
where $\ket{x}$ represent the computational basis states of $n$ qubits and 
$f(x)$ is the $\{0,1\}^{n}\to \{0,1\}$ Boolean function that needs to be 
evaluated. 
Notice that $(-1)^{f(x)}=\pm 1$ is just a real  phase factor.

The above states are achieved by starting from the equally weighted 
superposition of all possible $2^n$ states in the computational basis, 
namely the state

\begin{equation}
\ket{\psi_0}\equiv \frac{1}{\sqrt{2^n}}\sum_{x=0}^{2^n-1}\ket{x}\;,
\label{psi0}
\end{equation}
and then applying a unitary transformation $U_f$ which involves 
an additional qubit (a target qubit) and acts as follows

\begin{equation}
U_f\ket{x}\ket{y}=\ket{x}\ket{f(x)\oplus y} \;.
\label{Uf}
\end{equation}
In the above expression $\ket{y}$ is the state of the target 
qubit and $\oplus$ denotes addition 
modulo 2. 
The target qubit is initially prepared in the state 
$\ket{-}\equiv(\ket{0}-\ket{1})/\sqrt{2}$. In this way the target state is 
left unchanged by the action of $U_f$ and the state of the $n$-qubit register takes the form (\ref{psif}).
We will now discuss the entanglement properties of the multi-qubit 
``real equally weighted states'' $\ket{\psi_f}$. 

In the Deutsch-Jozsa algorithm the function $f$ to be evaluated is promised 
to be either constant or balanced (balanced means that it takes output 0 for 
half of the inputs and output 1 for the others). 
The state (\ref{psif}), where a minus sign is present in front of half of the
computational basis states and will therefore be called balanced state, 
can  be either separable or entangled. 
The state (\ref{psif}) is separable if and only if it can be expressed in the form

\begin{equation}
\ket{\psi_{sep}}\equiv \frac{1}{\sqrt{2^n}}\sum_{x=0}^{2^n-1}(-1)^{a\cdot x}
\ket{x}\;,
\label{psis}
\end{equation}
where $a$ is an $n$-bit string and $a\cdot x\equiv a_1x_1\oplus a_2x_2
\oplus...a_nx_n$. Therefore, apart from global phase factors, 
the number of dinstinct separable states of the form (\ref{psif}) is $2^n$.
If the function $f$ is constant the state
$\ket{\psi_f}$ is clearly separable, and apart from a global phase factor, 
it is the state $\ket{\psi_0}$ (in this case the string $a$ in the form 
(\ref{psis}) has all bits 0). 
Notice that the above state in Eq. (\ref{psis}) corresponds exactly 
to the state computed in the Bernstein-Vazirani problem \cite{bv}, where the 
function is promised to be of the form $f_a(x)=a\cdot x$ and the question is 
to find $a$. Therefore, for the Bernstein-Vazirani
problem, which can be solved as the Deutsch-Jozsa problem restricted to the 
particular class of functions of the form $f_a$, the qubits involved 
are never entangled. This has already been observed in \cite{meyer}.

In general, if the function $f$ is balanced the state 
(\ref{psif}) can then be 
either separable or entangled. We will now evaluate the number of separable 
versus entangled states. 
The number of possible balanced functions is given by
$N_{bal}=B(2^n,2^{n-1})$, where $B$ denotes the binomial coefficient.
The number of balanced functions corresponding to separable states, which 
are then necessarily of the form (\ref{psis}),
is $N_{bal,sep}=2(2^n-1)$. This value is obtained by noticing 
that two different balanced functions correspond to each $a$ (they would 
correspond to states $\ket{\psi_s}$ and $-\ket{\psi_s}$, which differ just 
in a global phase factor), while
the case of vanishing $a$ does not correspond to a balanced function and 
therefore must be subtracted.
The fraction of balanced functions that involve separable 
states with respect to the total is therefore 
given by $N_{bal,sep}/N_{bal} =2 (2^n-1)(2^{n-1}!)^2/2^n!$. 
Using the Stirling approximation, i.e. 
$\lim_{x\rightarrow\infty} x!\approx \sqrt{2\pi x}x^x e^{-x}$,
we arrive at the asymptotic limit
$\lim_{n\rightarrow\infty}N_{bal,sep}/N_{bal} \approx 
\sqrt{2\pi}(2^n-1)2^{\frac{n}{2}}/2^{2^n}$.
Thus, for large numbers of qubits 
the fraction of separable balanced states 
becomes exponentially small, i.e. for most balanced functions the 
Deutsch-Jozsa algorithm employs entangled states. 

In the following we will denote by $S_q$ the set of pure $q$-separable states,
i.e. states that can be written as tensor products of pure states of $q$
subsystems \cite{H3}.
So far we have evaluated
the fraction of fully separable states (i.e. $\in S_n$). 
We can also evaluate 
the fraction of states that are not fully separable but contain entanglement 
only between two qubits (i.e. $\in S_{n-1}\backslash S_n$). 
Actually, this kind of states are of the form $\ket{\psi_{sep,n-2}}\otimes
\ket{\psi_{ent,2}}$. There are $2(2^{n-2}-1)*8$ dinstinct balanced functions
which correspond to states of this form: the factor $2(2^{n-2}-1)$ is the 
number of balanced functions corresponding to fully separable states of $n-2$
qubits, while 8 is the number of entangled real equally weighted 
states of two qubits
(four states with just a minus sign in the superposition and four states with 
three minus signs). States of this type can occur for $B(n,2)$ different 
partitions of the $n$ qubits, therefore  the total number of balanced 
functions corresponding to states in $S_{n-1}\backslash S_n$ is 
$2(2^{n-2}-1)*8*B(n,2)$.
By taking the limit of large $n$ we can see that the fraction of these 
functions over the total number $N_{bal}$ is still exponentially small.

Rather than evaluating all possible classes of states we will answer the 
question whether most states in the Deutsch-Jozsa algorithm 
(in the limit $n\rightarrow\infty$)
are genuinely multipartite entangled. To this end,
we count the number of biseparable balanced states, i.e. states 
in $S_2$ corresponding to balanced functions that are of the form
\begin{equation}
\ket{\psi_{bisep}}=\ket{\psi_k}\otimes\ket{\psi_{n-k}},
\label{biseparable}
\end{equation}
and their fraction within
all balanced states. 
Here, it does not matter whether the constituting states are entangled or separable,
because any pure state that cannot be written in a biseparable form is fully entangled.
Before counting the number of biseparable balanced states, we 
will establish when a biseparable state is balanced.

{\bf Lemma:} { For a pure real equally weighted state of $n$ qubits which 
is $q$-separable, i.e.
\begin{equation}
\ket{\psi_{q-sep}}=\ket{\psi_{k_1}}\otimes\ket{\psi_{k_2}}\otimes ...
\ket{\psi_{k_p}}, \ \ \text{with} \ \ \sum_{i=1}^q k_i = n; \ q>1 \ ,
\end{equation}
where each $\ket{\psi_{k_i}}$ is real equally weighted, 
the state $\ket{\psi_{q-sep}}$ is
{\em balanced} if and only if at least one $\ket{\psi_{k_i}}$ is balanced.

{\em Proof:} Denote by $N_{+(-)}^{(k)}$ the number of plus (minus) signs of a
real equally weighted state with $k$ qubits. \\
"$\Rightarrow$": For a balanced $n$-qubit state we have $N_+^{(n)}=N_-^{(n)}$. 
For each partition with $k_i$ qubits in one subsystem, and 
$n-k_i$ in the other subsystem, we thus have 
$N_+^{(k_i)}\cdot N_+^{(n-k_i)}+ N_-^{(k_i)}\cdot N_-^{(n-k_i)}
=N_+^{(k_i)}\cdot N_-^{(n-k_i)}+ N_-^{(k_i)}\cdot N_+^{(n-k_i)}$
 or
$N_+^{(n-k_i)}(N_-^{(k_i)}-N_+^{(k_i)}) = 
N_-^{(n-k_i)}(N_-^{(k_i)}-N_+^{(k_i)})$.
The solution of this equation is either $N_+^{(n-k_i)}=N_-^{(n-k_i)}$ or
$N_+^{(k_i)}=N_-^{(k_i)}$, i.e. at least one of the two subsystems is in 
a balanced state. This argument holds  for all possible partitions. \\
"$\Leftarrow$": Assume without loss of generality that 
$\ket{\psi_{k_1}}$ is balanced, i.e. $N_+^{(k_1)}=N_-^{(k_1)}$.
The number of plus signs in the $n$-qubit state is 
$N_+^{(n)}=N_+^{(k_1)}\cdot N_+^{(n-k_1)}+ N_-^{(k_1)}\cdot N_-^{(n-k_1)}$,
and the number of total minus signs is 
$N_-^{(n)}=N_+^{(k_1)}\cdot N_-^{(n-k_1)}+ N_-^{(k_1)}\cdot N_+^{(n-k_1)}$.
Due to $N_+^{(k_1)}=N_-^{(k_1)}$ we find
$N_+^{(n)}=N_-^{(n)}$.
\hfill $\Box$

To count all biseparable balanced states we first
fix $k$, and also fix the partition in Eq. (\ref{biseparable}). The number
of real equally weighted states where at least one of the two subsystems is 
balanced
is $N_{bisep}(k)=B(2^k,2^{k-1})\cdot 2^{2^{n-k}}
 + B(2^{n-k},2^{{n-k-1}})\cdot 2^{2^k}
-B(2^k,2^{k-1})\cdot B(2^{n-k},2^{n-k-1})$.
This expression is derived by counting the number of balanced functions in the 
left term times the number of all functions in the right, plus vice versa,
minus the terms where both parts are balanced and we have already included
it before. 
Next, we have to sum over all possible bipartitions for fixed $k$ (which
leads to the binomial $B(n,k)$ as factor) and then have to sum over all $k$.
The only partition where this argument does not hold is the case
of $k=n/2$ for even $n$. Here the factor 1/2 is needed to ensure that 
the partitions are not counted twice.
Thus, the number of biseparable balanced states is given by

\begin{eqnarray}
N^{DJ}_{bisep}&=&\sum_{k=1}^{\lfloor (n-1)/2 \rfloor} B(n,k)N_{bisep}(k)\\
&+&\frac{1}{2}B(n,n/2)N_{bisep}(n/2)\delta_{n/2-\lfloor
n/2 \rfloor,0}  \ .\nonumber
\end{eqnarray}
We now want to find $\lim_{n\rightarrow\infty} N^{DJ}_{bisep}/N_{bal}$.
To this end, it is convenient to  write the above expression in the following
form

\begin{eqnarray}
N^{DJ}_{bisep}&=&\sum_{k=1}^{n-1} B(n,k)(B(2^k,2^{k-1})\cdot 2^{2^{n-k}}\\
&-&\frac{1}{2}B(2^k,2^{k-1})\cdot B(2^{n-k},2^{n-k-1}))\;.\nonumber 
\end{eqnarray}
For $n\to\infty$ the largest term in the above equation is the one with
$k=1$, and therefore we find that

\begin{equation}
\lim_{n\rightarrow\infty} N^{DJ}_{bisep}\leq 
\lim_{n\rightarrow\infty} 2 (n-1) n \cdot 2^{2^{n-1}}  \ .
\end{equation}
Using this upper bound, we arrive at
\begin{eqnarray}
\lim_{n\rightarrow\infty} \frac{N^{DJ}_{bisep}}{N_{bal}}& \leq & 
\lim_{n\rightarrow\infty} 2 (n-1)n \cdot 2^{2^{n-1}}/B(2^n,2^{n-1}) 
\nonumber \\
& =& \lim_{n\rightarrow\infty}
\frac{\sqrt{2\pi} \cdot n^2\cdot 2^{n/2}}{2^{2^{n-1}}} \ .
\end{eqnarray}
Thus, in the limit of $n\rightarrow\infty$ the number of biseparable
states among the balanced ones goes to zero, and we conclude that
for large $n$ the Deutsch-Jozsa algorithm typically employs 
genuine multipartite entanglement.

We will now discuss the case of Grover's algorithm. 
The state (\ref{psif}) is achieved after the first application of the 
oracle. In this case the function $f$ has output 1 for entries $x$ that 
correspond to solutions of the search problem and output 0 for values of $x$ 
that are not solutions. 
Let us denote with $M$ the number of solutions, which typically
is much smaller than the total number of entries $2^n$. 
The number of possible states of the form $\ket{\psi_f}$ is $N_M=B(2^n,M)$.

In the case of a single solution $M=1$ the state
(\ref{psif}) corresponds to an equally weighted superposition of all 
possible computational basis states with the same relative phases, 
except for a single term, which has relative phase -1 with respect to the 
others. Such a state is 
fully entangled, because a biseparable state would contain at least two minus
signs. Therefore, it is genuine multipartite 
entangled. The number of possible states with $M=1$
is clearly $2^n$, because the -1 phase can be in front of any of the $2^n$
computational basis states.

In general, we remember that a  necessary condition for the 
state (\ref{psif}) to be fully separable (i.e. of the form (\ref{psis})), is 
that the output of $f$ is constant or balanced. Therefore, for non-balanced 
or constant functions, namely whenever $M\neq 0,2^{n-1}$, the state 
(\ref{psif}) is always entangled.
Moreover, in the case of odd $M$ it is never possible to 
write the state as a tensor product because this would lead to an even number 
of -1 relative phases in the state $\ket{\psi_f}$. 
Therefore, whenever the number of solutions $M$ is odd the state
$\ket{\psi_f}$ is fully entangled.

Let us now consider the case $M=2$. 
Here the state (\ref{psif}) can be either biseparable or 
entangled. 
Actually, the only possibility to write it as tensor product
would be in the form

\begin{equation}
\ket{\psi_{bisep}}\equiv \frac{1}{\sqrt{2}}(\ket{0}+\ket{1})
\otimes\ket{\psi_{ent,n-1}}\;,
\label{psi2s}
\end{equation}
where $\ket{\psi_{ent,n-1}}$ represents a fully entangled state of $n-1$ 
qubits, of the form $(\ref{psif})$ and with just one  relative phase -1 in 
the superposition, and the partition is arbitrary. 
If the state cannot be written in this form in any partition, then it is 
fully entangled.   
As mentioned above, the total number of possible states with $M=2$ is 
$B(2^n,2)$, while the number of biseparable states among these is $n 2^{n-1}$.
Therefore, the fraction of biseparable states for $M=2$ is exponentially small
in the asymptotic limit. In other words, for large $n$ and $M=2$ the 
Grover algorithm typically employs genuine multipartite entanglement.

Let us now consider the case $M=4$. The state $\ket{\psi_f}$ in this case 
is biseparable, triseparable or fully entangled. 
Actually, it could be
factorised in the triseparable form 
\begin{equation}
\ket{\psi_{3-sep}}\equiv \frac{1}{{2}}(\ket{0}+\ket{1})^{\otimes 2}
\otimes\ket{\psi_{en,n-2}}\;,
\label{psi3s}
\end{equation}
where $\ket{\psi_{ent,n-2}}$ 
represents a fully entangled state of $n-2$ qubits,
of the form $(\ref{psif})$ and with just one  relative phase -1 in the 
superposition, and the partition is again arbitrary. The number of possible 
triseparable states of the above form is $2^{n-2}B(n,2)$.
The only other possibility to factor it is in the biseparable form 
(\ref{psi2s}), where 
$\ket{\psi_{ent,n-1}}$ represents a fully entangled state of $n-1$ qubits, 
of the form $(\ref{psif})$ and with two components with -1 relative phases in 
the superposition. 
If the state cannot be written in the above biseparable form or in the
triseparable form (\ref{psi3s}) for any partition, then it is fully entangled. 

The total number of biseparable states for $M=4$ (written as tensor product 
either in the triseparable form (\ref{psi3s}) or in the form (\ref{psi2s})) is 
$N^G_{bisep}=nB(2^{n-1},2)$.
The fraction of biseparable states is then given by

\begin{equation}
\lim_{n\rightarrow\infty} \frac{N^G_{bisep}}{N_{M=4}}
=\lim_{n\rightarrow\infty} \frac{nB(2^{n-1},2)}{B(2^{n},4)}\simeq 0\;, 
\end{equation}
namely the states employed are typically genuine multipartite entangled.

The above argument can be generalised to values of $M$ which are powers of 2:
for $M=2^k$, the initial state $\ket{\psi_f}$ can be biseparable, 
triseparable, ... $(k+1)$-separable or fully entangled.
When the state is $k+1$-separable it has to be of the form

\begin{equation}
\ket{\psi_{k+1-sep}}\equiv \frac{1}{{2^{k/2}}}(\ket{0}+\ket{1})^{\otimes k}
\otimes\ket{\psi_{ent,n-k}}\;,
\label{psik+1s}
\end{equation}
where $\ket{\psi_{ent,n-k}}$ is a fully entangled state of $n-k$ qubits with
a single term in the superposition having relative phase -1.
The reason for this structure is as follows: assume that the $k$ parties
would contain entanglement; then the number $M$ of solutions,
according to the Lemma, would be at least $M\geq 2^{n/2}$, while we require
$M\ll 2^n$. 
The number of states of the form (\ref{psik+1s}) is $2^{n-k}B(n,k)$.

When the state is $k$-separable it has to be of the form

\begin{equation}
\ket{\psi_{k-sep}}
\equiv \frac{1}{{2^{(k-1)/2}}}(\ket{0}+\ket{1})^{\otimes k-1}
\otimes\ket{\psi_{ent,n-k+1}}\;,
\label{psiks}
\end{equation}
where now $\ket{\psi_{ent,n-k+1}}$ is a fully entangled state of $n-k+1$ qubits
with two components having -1 relative phase. 
In general, for $M=2^k$, the state $\ket{\psi_f}$ can be $j$-separable, with 
$j\leq k+1$, or fully entangled.

For a generic even $M=2^q(2p+1)$, the state $\ket{\psi_f}$ can be always 
biseparable or fully entangled. In particular, the state can then be 
$j$-separable, with all values of $j$ ranging from 2 to $q+1$, or fully 
entangled. 
The number of biseparable states for even $M$ is in general given by 
$N^G_{bisep}(M)=nB(2^{n-1},M/2)$.
Therefore, the fraction of biseparable states is given by
\begin{equation}
\lim_{n\rightarrow\infty} \frac{N^G_{bisep}(M)}{N_{M}}
=\lim_{n\rightarrow\infty} \frac{nB(2^{n-1},M/2)}
{B(2^{n},M)}\simeq 0\;, 
\end{equation}
where we consider $q$ and $p$ finite and fixed.
We can then conclude that for any $M\ll N$ (as mentioned above, 
in our analysis it is sufficient 
that $M<2^{n/2}$) the states are typically multipartite entangled.

We will now consider Simon's algorithm \cite{simon}. In this case 
the function to be evaluated is promised to be a 
$\{0,1\}^n \to \{0,1\}^n$ periodic $2\to 1$ function and the task is to 
evaluate the period $r$ ($r$ is an $n$-bit string, with $r\neq 0$) with the 
smallest number of evaluations.
In other words, in Simon's case $f(x)=f(y)$ if and only if $x=y\oplus r$.
The first register, as in the previous cases, is composed of $n$ qubits
prepared in state $\ket{\psi_0}$, while the target register is now composed of 
another set of $n$ qubits, which are all prepared in state $\ket{0}$.

After the function evaluation step (\ref{Uf}), the global state of the two 
registers takes the form

\begin{equation}
\ket{\psi_{Simon}^{(2)}}\equiv\frac{1}{\sqrt{2^{n-1}}}\sum_{i=1}^{2^{n-1}}
(\ket{x_i}+\ket{x_i+r})\ket{f(x_i)}\;,
\label{psiSimon}
\end{equation}
where $x_i$ is the set of $2^{n-1}$ input values leading to different outputs.
Notice that the above state is always entangled between the two registers. 
The next step is a measurement 
on the second register, after which the state of the first register takes the
simple form

\begin{equation}
\ket{\psi_{Simon}}\equiv\frac{1}{\sqrt{2}}(\ket{\bar x}+\ket{\bar x+r})\;,
\label{psiSimon1}
\end{equation}
where $\bar x$ is now a random value among the $x_i$'s.
We will now study the entanglement properties of the states of the form
(\ref{psiSimon1}). Notice that, by keeping $r$ fixed and changing $\bar x$
in the above form, the entanglement properties of the state do not change 
because by applying just local operations (of the type $\sigma_x$) we can 
reach all the states of the above form (with different $\bar x$ and same $r$). 
We can then for simplicity analyse 
the properties of these states by choosing the particular value $\bar x=0$, 
namely

\begin{equation}
\ket{\psi_{Simon,0}}\equiv\frac{1}{\sqrt{2}}(\ket{0}+\ket{r})\;.
\label{psiSimon0}
\end{equation}
 
Notice that the number of possible different states of the above form,
corresponding to all the possible non vanishing values of $r$, is $2^n-1$.
If $r$ contains a single value 1, i.e. $wt(r)=1$ ($wt(r)$ is the weight of the 
binary string $r$, namely the number of 1's that it contains) then the state 
$\ket{\psi_{Simon,0}}$ is 
fully separable ($\in S_n$), because it is of the form $1/\sqrt{2}
(\ket{0}+\ket{1})\ket{0}...\ket{0}$. 
There are $n$ possible states of this type, corresponding to the different
strings of length $n$ with weight 1.

If $wt(r)=2$, then the state (\ref{psiSimon0}) 
contains two-qubits entanglement 
($\in S_{n-1}$). The number of states of this form is $B(n,2)$. 
This simple counting argument can be generalised to the case of an arbitrary 
weight for $r$: if  $wt(r)=k$ 
the state (\ref{psiSimon0})
belongs to $S_{n-k+1}$ and there are $B(n,k)$ of them.
The case of maximum weight is $wt(k)=n$ and then the state is 
$n$-partite entangled, of the GHZ form.
Notice that the most populated class of states corresponds to values
of the period $r$ with $wt(r)=
\lfloor n/2 \rfloor$, i.e. typically half of the qubits are entangled
with each other.

\vspace*{1cm}
In summary, we have elucidated the role of multipartite entanglement in the
Deutsch-Jozsa, the Grover and the Simon algorithms. 
For the 
Deutsch-Jozsa algorithm, multipartite entanglement within the first register
is needed to accommodate all possible (balanced) functions.
We have shown that the fraction of balanced functions corresponding to 
biseparable
states decreases exponentially with the number of qubits. In this sense most 
balanced functions involve genuine multipartite entanglement. 
In the Grover
algorithm, we have shown that the entanglement properties of
the initial state of the first register depend on the number of solutions
and we have demonstrated that such a state is also typically 
multipartite entangled, when a small number of items is searched for. 
In Simon's algorithm the situation is slightly different: in general the
first register and the target register are entangled. Moreover, the kind
of entanglement within the first register after the measurement of the target 
register depends on the weight of the period $r$: for increasing weight
it involves an increasing number of qubits. For large $n$, the states
employed typically involve entanglement among half of the constituent qubits.
We have thus shown that multipartite entanglement is an essential property in
the considered quantum algorithms.  

This work was supported in part by the EU project CORNER and by Deutsche Forschungsgemeinschaft
 (DFG).

\end{document}